\documentclass[%
 amsmath,amssymb,
 aps,pra,twocolumn,superscriptaddress
]{revtex4-1} 
\usepackage{braket}
\usepackage{graphicx}
\usepackage{dcolumn}
\usepackage{bm}
\usepackage{subfigure} 
\usepackage{hyperref,color}
\definecolor{darkblue}{rgb}{0.0,0.0,0.7}
\hypersetup{colorlinks,breaklinks,linkcolor=darkblue,urlcolor=darkblue,anchorcolor=darkblue,citecolor=darkblue}
\usepackage{physics}
\providecommand{\ha}{{\hat{a}}}
\providecommand{\had}{{\hat{a}^{\dagger}}}

\providecommand{\hb}{{\hat{b}}}
\providecommand{\hbd}{{\hat{b}^{\dagger}}}
\providecommand{\db}{{\delta\hat{b}}}

\providecommand{\hc}{{\hat{c}}}
\providecommand{\hcd}{{\hat{c}^{\dagger}}}
\providecommand{\dc}{{\delta\hat{c}}}
\providecommand{\dcd}{{\delta\hat{c}^{\dagger}}}
\providecommand{\hq}{{\hat{q}}}
\providecommand{\hp}{{\hat{p}}}
\providecommand{\hx}{{\hat{X}}}
\providecommand{\hy}{{\hat{Y}}}
\providecommand{\hH}{{\hat{H}}}
\providecommand{\hQ}{{\hat{Q}}}
\providecommand{\hP}{{\hat{P}}}

\begin{document}
\title{Optomechanical approach to controlling the temperature and chemical potential of light}

\author{Chiao-Hsuan Wang}
\affiliation{Joint Quantum Institute, College Park, MD 20742}
\affiliation{Department of Physics, University of Maryland, College Park, MD 20742}
\affiliation{Joint Center for Quantum Information and Computer Science, College Park, MD 20742}
\author{Jacob M. Taylor}
\affiliation{Joint Quantum Institute, College Park, MD 20742}
\affiliation{Joint Center for Quantum Information and Computer Science, College Park, MD 20742}
\affiliation{National Institute
of Standards and Technology, Gaithersburg, MD 20899}


\begin{abstract}
Massless particles, including photons, are not governed by particle conservation law during their typical interaction with matter even at low energies, and thus have no chemical potential. However, in driven systems, near equilibrium dynamics can lead to equilibration of photons with a finite number, describable using an effective chemical potential [M. Hafezi \textit{et al.}, Phys. Rev. B \textbf{92}, 174305 (2015)]. Here we build upon this general concept with an implementation appropriate for a photon-based quantum simulator.
We consider how laser cooling of a well-isolated mechanical mode can provide an effective low-frequency bath for the quantum simulator system. We show that the use of auxiliary photon modes, coupled by the mechanical system, enables control of both the chemical potential and temperature of the resulting photonic quantum simulator's grand canonical ensemble.
\end{abstract}

\pacs{}
\maketitle

\section{Introduction}

Massless particles, including photons, usually do not exhibit number-conserving interactions and thus are described in equilibrium by the canonical ensemble -- they have no chemical potential. One of the consequences of a vanishing chemical potential is that for a cold photon gas in thermal equilibrium, rather than forming a condensate as massive bosons do, the occupation number of photons drops significantly as described by the theory of black body radiation. Thus, the vacuum is the typical ground state of such systems. As a promising avenue for quantum simulation \cite{Aspuru-Guzik2012,Noh2017,hartmann2016}, quantum photonic simulators can be created with tunable coupling and interactions. Typically, they are driven far from equilibrium by a laser or coherent microwave source to populate a sufficient number of photons to create an interesting many-body state of light. Simple examples include observations of quasiequilibrium Bose-Einstein condensate of photons in recent experiments using cavity polaritons \cite{Kasprzak2006,Balili2007,Deng2010,Tosi2012} or with dye microcavities \cite{Klaers2010a} under pumping and loss process, and self-organization of atoms and open Dicke model phase transition in the setting of cavity quantum electrodynamics \cite{Black2003,Dimer2007,Nagy2008,Baumann2010}. More complex versions now include driven arrays of Josephson-junction-based devices \cite{Fitzpatrick2017}.

Current experiments either use  approximate number conservation \cite{Weitz2013,Ma2017} (enabling an effective chemical potential description with a phenomenological temperature) or are far-from-equilibrium and instead described by a steady state. However, it has been recently proposed that in parametrically driven systems, near equilibrium dynamics can lead to equilibration of photons into a thermodynamic ensemble with a finite number of photons \cite{Hafezi2015}. This Gibbs-like ensemble then has an effective chemical potential, and the dynamics admits a near-equilibrium description without solving the full driven-dissipative non-equilibrium quantum problem \cite{Wang2016}. The key challenge for this approach is introducing an appropriate bath to the photonic system via a parametric coupling.

A canonical parametric process is the generation of sidebands of light by the motion of a mirror. Studies of the interaction between light and motion have paved the way for preparation and manipulation of non-classical states of light and macroscopic mechanical resonators \cite{kippenberg2008,Marquardt2009,Clerk2010,aspelmeyer2014}.  The development and advancement of optomechanical cooling techniques \cite{marquardt2007,wilson-rae2007,Marquardt2008,elste2009}, recently reaching the quantum backaction limit \cite{peterson2016}, have made possible the preparation of the quantum ground state of a mechanical resonator \cite{oconnell2010,chan2011,teufel2011}, squeezed states of light \cite{rabl2004,purdy2013,Safavi-Naeini2013a}, realization of nonlinear optics \cite{nunnenkamp2011,rabl2011,borkje2013,lemonde2013,xu2015,lemonde2016}, and bath engineering for photons using the mechanical degree of freedom \cite{toth2017} in optomechanical platforms.

Here we build upon the general concept of chemical potential in driven systems with an optomechanical implementation appropriate for a nonlinear photonic or microwave quantum simulator, taking full advantage of the advances in laser cooling and related techniques in optomechanics to control the effective bath for the photonic system. The parametric optomechanical interaction between the optical system and the low-frequency bath is provided through a beam-splitter coupling between the optical system and another laser-driven mode, which can be realized in a Michelson-Sagnac interferometer~ \cite{yamamoto2010,sawadsky2015}. The use of multiple photon modes enables control of both the chemical potential, by drive frequency, and temperature, by drive amplitude, of the resulting photonic grand canonical ensemble.

\section{Optomechanical Implementation of a parametric bath for photons}
Here we propose an optomechanical implementation for a controllable bath that leads to a grand canonical ensemble of photons with definite temperature and chemical potential by parametric coupling in a driven system. One natural candidate to engineer the parametric coupling is through optomechanics, where thermal (mechanical) excitation can create sideband photons from a pump laser, leading to an effective photonic bath. Consider a beam-splitter-type coupling, which is common in so-called mirror-in-the-middle systems \cite{jayich2008,xu2014,stambaugh2015} and can also be realized in Michaelson-Sagnac geometry \cite{yamamoto2010,sawadsky2015}, between optical modes $\ha$ and $\hb$ and the motion of the mechanical resonator $\hq$, $\hat{V}_{qa}= -\hbar  G_{a0}(\hbd \ha + \had \hb)\hq$, where $G_{a0}$ is the coupling parameter between $\ha$, $\hb$, and $\hq$. By driving the photonic mode $\hb$ with a strong laser of frequency $\nu_b$, we can  expand $\hb$ as a small quantum fluctuation $\db$ around a large steady-state mean value $b_s$, $\hb(t)=b_s e^{-i \nu_b t}+ \db(t)$, and the interaction can be linearized by neglecting the quantum fluctuations $\db(t)$,  
\begin{align} 
\hat{V}_{qa}(t) \approx -\hbar  G_{a0} b_s (\ha  e^{i \nu_b t}+ \had e^{-i \nu_b t})\hq.
\label{Hqc(t)}
\end{align} 
We assume that the coupling strength is weak compared to optical energies, $\hbar g_a \ll \hbar \nu_b$, which is typically true for optomechanical interactions. Here $g_a= G_{a0} b_s q_{\rm ZPF}$ is the pump enhanced coupling between $\hq$, $\ha$, and $\hb$ in the unit of frequency and $q_{\rm ZPF}=\sqrt{\frac{\hbar}{2 M \omega_m}}$ is the mechanical zero-point fluctuation. Given this weak parametric coupling and sufficiently small optical losses, one expects the system to reach an equilibrium state describable by a grand canonical ensemble with chemical potential $\hbar \nu_b$, as shown in Ref. \cite{Hafezi2015}.

\begin{figure}[htbp]
\begin{center}
\includegraphics[width=0.47 \textwidth]{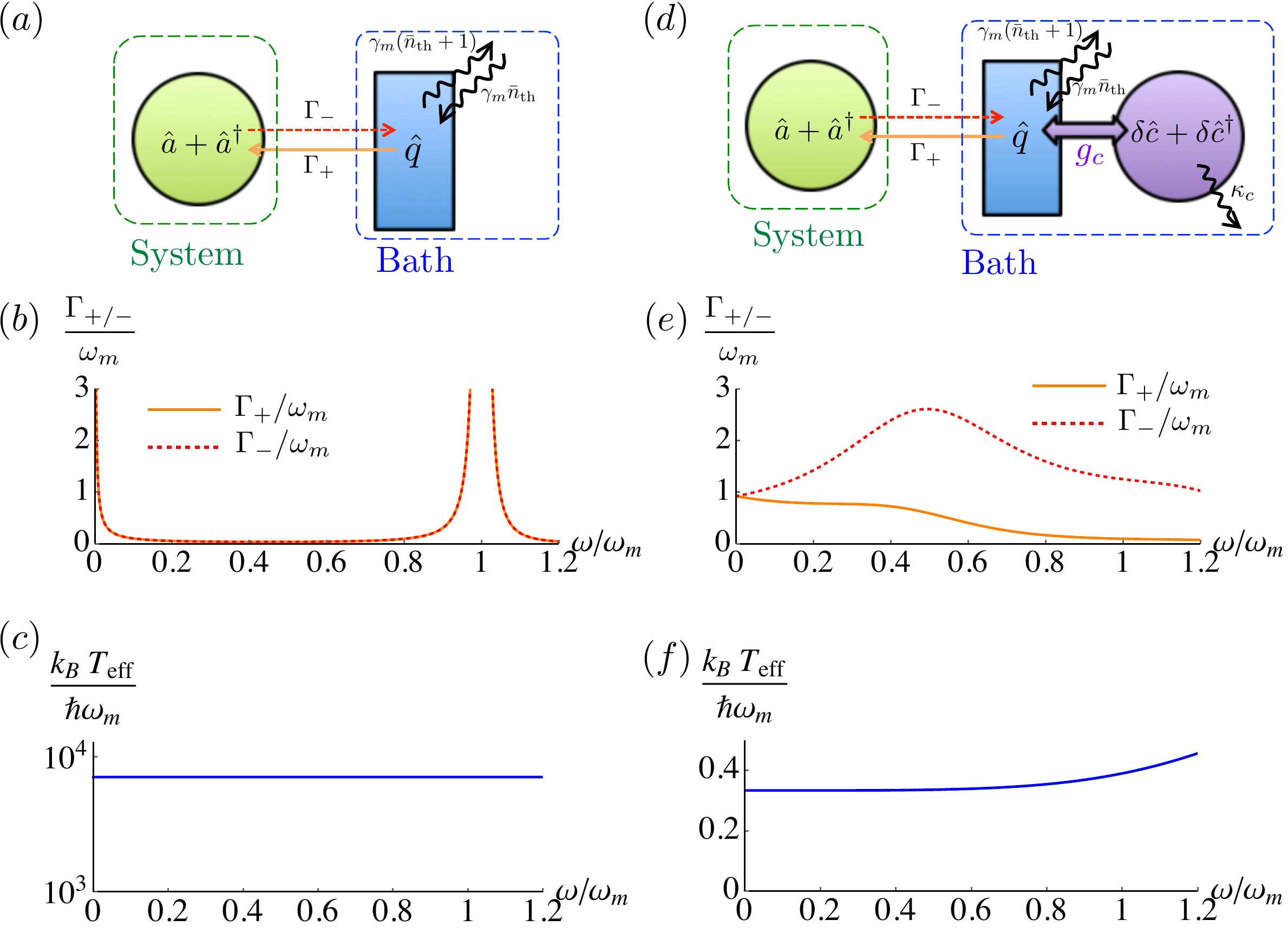}
\caption{Comparison between cases of (a)-(c) a mechanical resonator as a bath and (d)-(f) a laser-cooled mechanical resonator as a bath for the many-body photonic system. (a) Schematic diagram of a mechanical resonator $\hq$ serving as an effective bath for the many-body photonic system $\ha$ through optomechanical interaction. (b) Photon emission coefficient $\Gamma_+$ (orange solid line) and absorption coefficient $\Gamma_-$ (red dashed line) due to coupling with the mechanical resonator. The coefficients have an out-of-range peak centered at the mechanical resonant frequency $\omega_m$, with a small width characterized by the mechanical damping rate $\gamma_m$, leading to efficient thermalization only within the narrowband around $\omega_m$. (c) Effective bath temperature $k_B T_{\rm eff}$ of the mechanical resonator. (d) Schematic diagram of a mechanical resonator $\hq$, laser-cooled by a cooling mode $\hc$, serving as a thermal bath to the photonic many-body system $\ha$ through optomechanical interaction. (e) Photon emission coefficient $\Gamma_+$ (orange solid line) and absorption coefficient $\Gamma_-$ (red dashed line) due to coupling with the laser-cooled mechanical resonator, suggesting a much broader bandwidth towards low frequencies. (f) Effective bath temperature $k_B T_{\rm eff}$ of a laser-cooled mechanical resonator, which is lowered by a factor of $10^{-5}$ in comparison to the case without laser-cooling. These plots are generated with the parameters $\Delta_c=-\omega_m=-\sqrt{\frac{3}{4}}\kappa_c$, $\beta=10^{-4}\omega_m$, $\gamma_m=10^{-6}\omega_m$, the cavity-enhanced system-bath coupling $g_a=0.45\omega_m$, and the mechanical resonator-cooling cavity coupling $g_c=0.45 \omega_m$.}
\label{fig:Bath}
\end{center}
\end{figure}

In practice, there are three fundamental limits to this optomechanical approach. First, the response of a high quality factor resonator is narrow band, characterized by its mechanical damping rate, leading to thermalization for transitions only very near the mechanical resonances. Second, the mechanical temperature may be too high even in cryogenic settings, compared to, for example, the relevant photonic nonlinear terms around 100 MHz one may be using to implement a many-body Hamiltonian [see Fig.~\ref{fig:Bath}(a)-\ref{fig:Bath}(c)]. Third, the optomechanical interaction requires a strong pump field in mode $\hb$, which we would like to not pollute our many-body optical system $\ha$. That is, we want no steady-state coherence generated in our optical system $\ha$ by the pump ($a_s=0$).

\begin{figure}[htbp]
\begin{center}
\includegraphics[width=0.47 \textwidth]{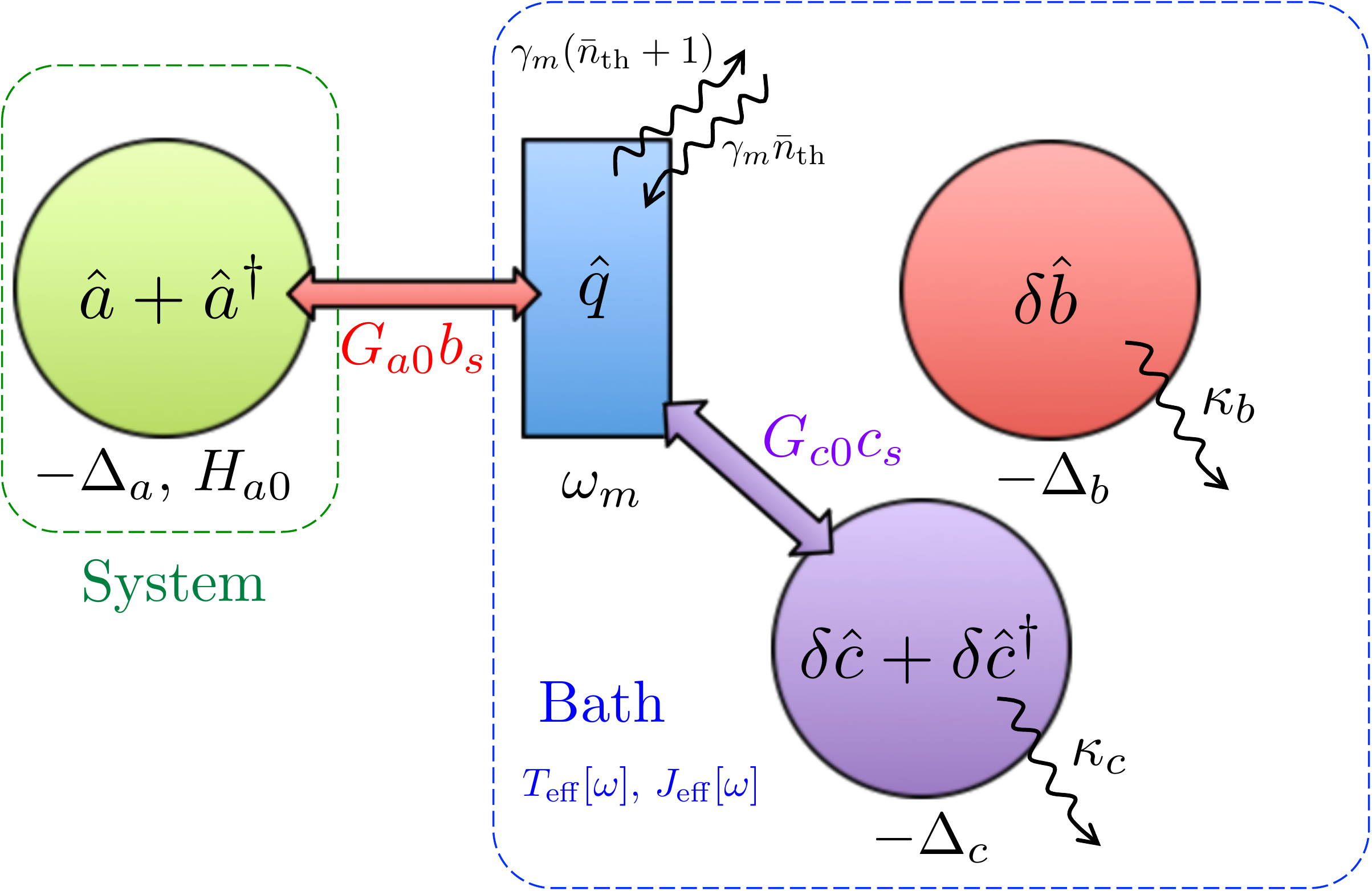}
\caption{Schematic diagram of the optomechanical implementation of a parametric bath for photons. Here, $\hb$, $\hc$, $\hq$, and their dissipative environments all together serve as a bath for the photonic system $\ha$ as described in the linearized Heisenberg-Langevin equations (\ref{LInOut}).}
\label{fig:Schematic}
\end{center}
\end{figure}

To conquer the first two challenges, we propose adding an additional optical cooling mode $\hc$ to broaden the mechanical bandwidth and lower the temperature via laser cooling [see Fig.~\ref{fig:Bath}(d)-\ref{fig:Bath}(f)]. Specifically, the quantum Brownian motion theory applies to the coupled three optical modes and one mechanical mode by treating $\hb$, $\hc$, $\hq$ all together as an effective heat bath for the system $\ha$ as we show below. The last problem we solve only technically via a potential experimental design using Michelson-Sagnac geometry \cite{yamamoto2010,sawadsky2015}, but the underlying concept of using beam-splitter optomechanical interactions for pump rejection should be extensible to many other configurations.

We now focus on our solutions to the first two challenges. Specifically, we consider how the correlation functions, which describe the full dynamics of the mechanical system, including its bath, are modified by laser cooling using, e.g., an additional photonic mode $\hc$. This allows us to connect the full non-Markovian theory that describes the original mechanical system to a new effective bath that includes the laser cooling and is close to Markovian. Thus the $\hq$ degree of freedom is effectively promoted to a bath operator $\hat{B}$ whose correlation functions mimic the desired bath (including chemical potential) for the many-body photonic system $\ha$. We will show how $\hb$, $\hc$, $\hq$, and their dissipative environments all together serve as a bath for the many-body system $\ha$ (see Fig.~\ref{fig:Schematic}). Specifically, the Hamiltonian reads
\begin{align}
\hH=& \hH_a+\hH_0+\hat{V}_{qa}+\hat{V}_{qc}+\hH_{\rm drive,b}+\hH_{\rm drive,c}\notag\\
&+\hH_{\kappa_b}+\hH_{\kappa_c}+\hH_{\gamma_m},
\label{optoH}
\end{align}
where $\hH_a$ is some general system Hamiltonian for $\ha$ and might contain nonlinear terms. We have
\begin{align}
\hH_0= \hbar\omega_a \had\ha+\hbar\omega_b \hbd \hb+\hbar\omega_c \hcd \hc+\frac{\hp^2}{2M}+\frac{M\omega_m^2(\hq-q_0)^2}{2},
\label{H0}
\end{align}
\begin{align}
\hat{V}_{qa}= -\hbar  G_{a0}(\hbd \ha + \had \hb)\hq,
\label{Hqc}
\end{align}
\begin{align}
\hat{V}_{qc}= -\hbar G_{c0} (\hcd \hc)\hq,
\label{Hqb}
\end{align}
\begin{align}
\hH_{\rm drive,b}= i \hbar \sqrt{\kappa_b} \left(\hbd \hb_{\rm in}e^{- i \nu_b t} - \hb \hbd_{\rm in}e^{ i \nu_b t}\right) ,
\label{Hda}
\end{align}
\begin{align}
H_{\rm drive,c}= i \hbar \sqrt{\kappa_c} \left(\hcd 
\hc_{\rm in}e^{- i \nu_c t} -  \hc 
\hcd_{\rm in}e^{ i \nu_c t}\right).
\label{Hdb}
\end{align}
Here $\omega_a$, $\omega_b$, and $\omega_c$ are the frequencies of the optical modes $\ha$, $\hb$, and $\hc$; $M$, $\omega_m$, and $\hp$ are the mass, mode frequency, and momentum of the mechanical resonator; $G_{c0}$ is the coupling parameter between the optical mode $\hc$ and the mechanical resonator $\hq$; $H_{\kappa_b}$, $H_{\kappa_c}$, and $H_{\gamma_m}$ are the dissipative interactions of the cavity modes $\hb$ and $\hc$ and mechanical modes $\hq$ with the environment with damping rates $\kappa_b$, $\kappa_c$, and $\gamma_m$ respectively. Note that we have assumed the perfect cavity limit, i.e., no internal losses inside the high quality factor cavities, such that the dissipation in cavity modes comes solely from the coupling with the external drive. The loss rate of the system $\ha$ is therefore zero, $\kappa_a=0$, since we are driving $\hb$ and $\hc$ modes only. The interaction $\hat{V}_{qc}$ between the mechanical mode and cooling mode $\hc$ can also be beam-splitter-like.

We now move to a rotating frame through $U=e^{i(\nu_b \had \ha+\nu_b \hbd \hb+\nu_c \hcd \hc)t}$. $H_a$ can be decomposed into $H_{a0}+H_{a,\perp}$. Here $H_{a0}$ is the particle-number conserving part (comprising of $\ha \had$ and $\had\ha$ pairs) while $H_{a,\perp}$ includes all terms that do not conserve the total number of particles. Assuming $||H_{a,\perp}|| \ll \hbar\nu_b$ and weak beam-splitter coupling $ \hbar G_{a0} q_{\rm ZPF}\ll \hbar\nu_b$, we can make the rotating wave approximation such that $H^r_{a}(t) =U(t) H_a U^{\dagger}(t)\approx H_{a0}$. The full quantum Langevin equations of motion based on the input-output formalism (following \cite{Paternostro2006}; see also \cite{Gardiner1985,Clerk2010}) now read
\begin{align}
\dot{\ha}=&\frac{i}{\hbar}\left[H_{a0},\ha \right] + i \Delta_a\ha + i G_{a0} \hq \hb,\notag\\
\dot{\hb}=&i(\Delta_a -\frac{\kappa_b}{2})\hb+i G_{a0} \hq \ha+\sqrt{\kappa_b}\hb_{\rm in}(t),\notag\\
\dot{\hc}=&i(\Delta_c -\frac{\kappa_c}{2} )\hc + i G_{c0} \hq\hc+\sqrt{\kappa_c}\hc_{\rm in}(t),\notag\\
\dot{\hq}=&\frac{\hp}{M},\notag\\
\dot{\hp}=&-M\omega_m^2 (\hq-q_0)-\gamma_m \hp+\hbar G_{a0} (\hbd \ha +\had \hb)\notag\\
&+\hbar G_{c0} \hcd \hc+\hat{F}_{\rm in}(t).
\label{InOut}
\end{align}
We have defined the optical detunings $\Delta_a=\nu_b-\omega_a$, $\Delta_b=\nu_b-\omega_b$, and $\Delta_c=\nu_c-\omega_c$. The cavity input fields $\hb_{\rm in}=b_{\rm in,s}+\db_{\rm in}$ and $\hc_{\rm in}=c_{\rm in,s}+\dc_{\rm in}$ have classical drive amplitudes $b_{\rm in,s}$ and $c_{\rm in,s}$ and quantum vacuum noise parts $\db_{\rm in}$ and $\dc_{\rm in}$, respectively, while the mechanical motion is affected by a Brownian stochastic force $\hat{F}_{\rm in}(t)$ \cite{Gardiner1985,Grabert1984,Hanggi2005,Weiss2008}. 
The baths are described by the noise correlators,
\begin{align}
&\left\langle \db_{\rm in}(t)\db^{\dagger}_{\rm in}(t{'})\right\rangle=\delta(t-t{'}),\notag\\
&\left\langle \db_{\rm in}(t)\db_{\rm in}(t{'})\right\rangle=\left\langle \db^{\dagger}_{\rm in}(t)\db_{\rm in}(t{'})\right\rangle=0,
\label{noisea}
\end{align}
\begin{align}
&\left\langle \dc_{\rm in}(t)\dc^{\dagger}_{\rm in}(t{'})\right\rangle=\delta(t-t{'}),\notag\\
&\left\langle \dc_{\rm in}(t)\dc_{\rm in}(t{'})\right\rangle=\left\langle \dc^{\dagger}_{\rm in}(t)\dc_{\rm in}(t{'})\right\rangle=0,
\label{noiseb}
\end{align}
\begin{align}
\left\langle \hat{F}_{\rm in}(t) \hat{F}_{\rm in}(t{'})\right\rangle =&\frac{\hbar}{\pi} \int_0^{\infty} \mathrm{d}\omega J[\omega] \{ e^{-i\omega(t-t{'})}(1+\bar{n}_{\rm th}[\omega])\notag\\
&+e^{i\omega(t-t{'})}\bar{n}_{\rm th}[\omega] \} \notag\\
=&\frac{\hbar}{\pi} \int_0^{\infty} \mathrm{d}\omega J[\omega]\left[ \coth \left(\frac{\beta \hbar \omega}{2}\right)\cos[\omega(t-t{'})]\right.\notag\\&
-\left.i\sin[\omega(t-t{'})]\right] .
\label{noiseq}
\end{align}
Here $\bar{n}_{\rm th}[\omega]=\frac{1}{e^{\beta \hbar \omega}-1}$ is the bosonic occupation function at thermal equilibrium for the mechanical thermal environment, and for specificity we assume the spectral density $J[\omega]= \gamma_m \omega M e^{-\omega/\omega_a}$ for Ohmic damping, though other baths work as well. In the end, the laser cooling changes the correlation functions sufficiently to eliminate these effects. However, putting in the Ohmic form of $J[\omega]$ now helps us connect this system to the effective spectral density $J_{\rm eff}[\omega]$ we obtain later by looking at the correlation functions of $\hq$ in the laser cooling regime.

Assuming that the laser fields are strong, we now separate the dynamics of the operators into their semiclassical steady state values and quantum fluctuations, $\hat{\mathcal{O}}=\mathcal{O}_s+\delta \hat{\mathcal{O}}$ and $\hat{\mathcal{O}}=\ha,\had,\hb,\hbd,\hc,\hcd,\hq,\hp$. We are interested in the case $\sqrt{\left\langle \had \ha \right\rangle}=0$ such that the occupation of the system is not driven by the laser pump. Note that we need to set $q_0=-\hbar G_{c0} \abs{c_s}^2/M\omega_m^2$ to balance the displacement induced by the constant radiation pressure force and make $q_s=0$ such that the vanishing steady state solution for the system $a_s=0$ is allowed in the coupled equation of motion. Solving for the steady state solution through Eq.~(\ref{InOut}) with the above condition, we have $b_s=\sqrt{\kappa_b} b_{\rm in,s}/(i\Delta_d-\kappa_b/2)$, $c_s=\sqrt{\kappa_c} c_{\rm in,s}/(i\Delta_c-\kappa_c/2)$, and $a_s=q_s=p_s=0$. We can take $b_s$ and $c_s$ to be real by absorbing the complex phase into the definition of the laser amplitudes. Since $a_s=q_s=p_s=0$, our system $\ha$ and the resonator mode $\hq$ are not displaced. 

By keeping only the linear terms in fluctuations, though making no assumption about $H_a$ for the many-body system of interest $\ha$, we arrive at the linearized Heisenberg-Langevin equations for the quantum dynamics:
\begin{align}
\dot{\ha}&=\frac{i}{\hbar}\left[H_{a0},\ha \right] + i \Delta_a\ha + i G_{a0} b_s \hq,\notag\\
\delta\dot{\hb}&=(i \Delta_b -\frac{\kappa_b}{2})\db+\sqrt{\kappa_b}\db_{\rm in}(t),\notag\\
\delta\dot{\hc}&=(i \Delta_c -\frac{\kappa_c}{2})\dc+iG_{c0} c_s \hq+\sqrt{\kappa_c}\dc_{\rm in}(t),\notag\\
\dot{\hq}&=\frac{\hp}{M},\notag\\
\dot{\hp}&=-M\omega_m^2 \hq - \gamma_m \hp +\hbar G_{a0} b_s (\ha+\had) \notag \\
&\ \ +\hbar G_{c0} c_s (\dc+\dcd)+ \hat{F}_{\rm in}(t).
\label{LInOut}
\end{align}
One can see that the optical field $\db$ now decouples from all the other modes, only entering the dynamics with its steady state value $b_s$ as an enhancement of the coupling between $\hq$ and $\ha$.

Note that one can see explicitly the form of a parametric coupling between $\ha$ and $\hq$ by rotating back to the laboratory frame of the system c, $\hat{V}_{qa}^{\rm lin.,lab}(t)=-\hbar G_{a0} b_s (\ha e^{i \nu_b t} + \had e^{-i \nu_b t}) \hq$. We stress that the original beam-splitter-type coupling between $\ha$ and $\hb$ is essential for the pump rejection purpose such that the classical amplitude of $\hb$ mediated the sinusoidal parametric coupling without pumping the system directly.

\section{Effective Bath Spectral Density and Temperature}
The system-bath coupling is proportional to $\ha+\ha^{\dagger}$, which will lead to a force-like term as in quantum Brownian motion. We show that the correlation function of the mechanical resonator can be expressed in a form analogous to Eq.~(\ref{noiseq}) as
\begin{align}
&C_{qq}(t)\equiv \left\langle \hq_I(t)\hq_I(0)\right\rangle\notag\\
&=\frac{\hbar}{\pi} \int_0^{\infty} \mathrm{d}\omega J_{\rm eff}[\omega]\left[ \coth \left(\frac{\hbar \omega\beta_{\rm eff}[\omega] }{2}\right)\cos(\omega t)-i\sin(\omega t)\right],
\label{Cqqopto}
\end{align}
where $\hq_I(t)=e^{iH_Bt}\hq e^{-iH_Bt}$ is the coordinate field in the interaction picture. The evolution of the many-body system $\ha$ can be described as Langevin equations \cite{Gardiner2004,Weiss2008}:
\begin{align}
\dot{\hat{X}}_a=&-\Delta_a \hy_a+\frac{i}{\hbar}\left[H_{a0},\hx_a \right],\notag\\
\dot{\hat{Y}}_a=&\Delta_a \hx_a+\frac{i}{\hbar}\left[H_{a0},\hy_a \right]+\hat{\xi}_{Y_a}(t)\notag\\
&-2 \hbar G_{a0}^2 b_s^2 \int_0^{t} \mathrm{d}t{'} \hx_a(t{'}) \frac{\mathrm{d}}{\mathrm{d}t}\gamma_{\rm eff}(t-t{'}),
\label{clangevin}
\end{align}
with a Langevin force-like term 
\begin{align}
\hat{\xi}_{Y_a}(t)=\sqrt{2}G_{a0} b_s \hq_I(t) .
\label{xiYa}
\end{align}
Here we have introduced cavity mode quadratures $\hx_a=\sqrt{\frac{1}{2}}(\ha+\had)$ and $ \hy_a=i\sqrt{\frac{1}{2}}(\had-\ha)$. The effective damping kernel is defined as $\gamma_{\rm eff}(t)=\Theta(t)\frac{2}{\pi}\int_0^{\infty}\frac{J_{\rm eff}[\omega]}{\omega}\cos(\omega t) \mathrm{d}\omega,$ and $\hq_I(t)$ determines the properties of the stochastic force $\hat{\xi}_{Y_a}(t)$.

To find the interaction operator $\hq_I(t)$, it is equivalent to solve a set of coupled equations of motion for $\dc$ and $\hq$ as in Eq.~(\ref{LInOut}) but without the system-bath coupling terms. Solving the equations in Fourier domain $\tilde{\mathcal{O}}[\omega]=\int_{-\infty}^{\infty} \mathrm{d}t e^{i\omega t}\delta\hat{\mathcal{O}}_I(t)$ and defining the pump-enhanced coupling $G_c=G_{c0} c_s$, we now have
\begin{align}
-i\omega\tilde{c}[\omega]=&(i \Delta_c -\frac{\kappa_c}{2})\tilde{c}[\omega]+\sqrt{\kappa_c}\tilde{c}_{\rm in}[\omega]+iG_c \tilde{q}[\omega],\notag\\
-i\omega\tilde{c}^{\dagger}[\omega]=&(-i \Delta_c -\frac{\kappa_c}{2})\tilde{c}^{\dagger}[\omega]+\sqrt{\kappa_c}\tilde{c}^{\dagger}_{\rm in}[\omega]-iG_c \tilde{q}[\omega],\notag\\
-i\omega\tilde{q}[\omega]=&\frac{\tilde{p}[\omega]}{M},\notag\\
-i\omega\tilde{p}[\omega]=&-M\omega_m^2\tilde{q}[\omega]-\gamma_m \tilde{p}[\omega]+\tilde{F}_{\rm in}[\omega]
\notag\\&+\hbar G_c (\tilde{c}[\omega]+\tilde{c}^{\dagger}[\omega]).
\label{FourierEOM}
\end{align}
We define the bare mechanical susceptibility
\begin{align}
\chi_{q,0}^{-1}[\omega]=-M\omega^2+M\omega_m^2-iM\omega\gamma_m,
\end{align}
and the optomechanical modification from the cooling mode $\hc$,
\begin{align}
\Sigma[\omega]=\hbar G_c^2 \left( \frac{1}{(\Delta_c+\omega)+i\kappa_c/2}+\frac{1}{(\Delta_c-\omega)-i\kappa_c/2}\right),
\end{align}
\begin{widetext}
such that $\chi_{q}^{-1}[\omega]=\chi_{q,0}^{-1}[\omega]+\Sigma[\omega]$. Note that $\chi_{q}[-\omega]=(\chi_q[\omega])^*$. We have
\begin{align}
\tilde{q}[\omega]=\chi_q[\omega]\left[\tilde{F}_{\rm in}[\omega]+\left( \frac{\hbar G_c \sqrt{\kappa_c}\tilde{c}_{\rm in}[\omega]}{-i(\Delta_c+\omega)+\kappa_c/2}+\frac{\hbar G_c \sqrt{\kappa_c}\tilde{c}_{\rm in}^{\dagger}[\omega]}{i(\Delta_c-\omega)+\kappa_c/2}\right) \right]=\chi_q[\omega](\tilde{F}_{\rm in}[\omega]+\tilde{C}_{\rm in}[\omega]).
\label{tildeqopto}
\end{align}
\end{widetext}
The position autocorrelation function thus has two contributions: $C_{qq}(t)=C_{qq,F}(t)+C_{qq,c}(t)$.
One contribution is from the Ohmic mechanical bath,
\begin{align}
C_{qq,F}(t)=&\int_{0}^{\infty}\mathrm{d}\omega \frac{\hbar J[\omega] \abs{\chi_q[\omega]}^2}{\pi} \notag\\& \times \left[ \coth \left(\frac{\beta \hbar \omega}{2}\right)\cos(\omega t)-i\sin(\omega t)\right],
\label{CqqF}
\end{align}
and the other is from the optical cooling and counter-rotating (heating) terms:
\begin{align}
C_{qq,c}(t)=\int_{0}^{\infty}\mathrm{d}\omega \hbar^2 G_c^2 \abs{\chi_q[\omega]}^2\left( e^{-i\omega t}L[\omega]+e^{i\omega t}L[-\omega]\right).
\label{CqqB}
\end{align}
Here $L[\omega]$ is a Lorentzian function of frequency centered at $-\Delta_c$ with a width $\kappa_c$,
\begin{align}
L[\omega] \equiv\frac{\kappa_c/2\pi}{(\omega+\Delta_c)^2+\kappa_c^2/4}.
\label{Lorentzian}
\end{align}

Compared with Eq.~(\ref{Cqqopto}), we arrive at a new quantum Brownian motion bath with a modified spectral density $J_{\rm eff}[\omega]$, a frequency-dependent temperature $T_{\rm eff}[\omega]$, and an in general non-Markovian damping kernel $\gamma_{\rm eff}(t)$. The effective spectral density is
\begin{align}
J_{\rm eff}[\omega]=\abs{\chi_q[\omega]}^2\left\{ J[\omega]+\pi\hbar G_c^2(L[\omega]-L[-\omega])\right\},
\label{jeff}
\end{align}
and the effective temperature $T_{\rm eff}[\omega]=1/k_B \beta_{\rm eff}[\omega]$ is given implicitly by
\begin{align}
\coth&\left(\frac{\beta_{\rm eff} [\omega]\hbar \omega}{2}\right)\notag\\=&\frac{J[\omega] \coth \left(\frac{\beta\hbar \omega}{2}\right)+\pi\hbar G_c^2 \left(L[\omega] +L[-\omega] \right)}{J[\omega]+\pi\hbar G_c^2 \left(L[\omega] -L[-\omega] \right)},
\label{cotheff}
\end{align}
or equivalently by a detailed-balance-like condition
\begin{align}
e^{\hbar \omega \beta_{\rm eff}[\omega]}=\frac{J[\omega] (\bar{n}[\omega] +1)+\pi\hbar G_c^2 L[\omega] }{J[\omega] \bar{n}[\omega] +\pi\hbar G_c^2 L[-\omega] }.
\label{detailbalance}
\end{align}
We note that working in the red-detuned regime such that $\Delta_c<0$, when $\frac{L[\omega]}{L[-\omega]} > e^{\hbar \omega \beta}$, we have $\beta_{\rm eff}[\omega]>\beta$, consistent with the cooling mechanism.

The effective temperature and spectral density determine the equilibrium distribution of the photons through a detailed-balance condition~\cite{Clerk2010}. Specifically, consider the quantum noise spectrum defined as
\begin{align}
S_{qq}[\omega]=\int_{- \infty}^{\infty} \mathrm{d}t e^{i \omega t } C_{qq}(t).
\label{Sxx}
\end{align}
If the $\ha$ mode is at a frequency $\Omega=-\Delta_a=\omega_a-\nu_b$, then the corresponding Fermi's golden rule transition rates of emitting one photon by absorbing energy from the effective bath $R_{\rm n \rightarrow n+1}[\Omega]$ and losing one photon to the effective bath $R_{\rm n \rightarrow n-1}[\Omega]$ can be expressed in terms of the quantum noise spectrum as 
\begin{align}
&R_{\rm n \rightarrow n+1}[\Omega]=(n+1)\frac{g_a^2}{q_{\rm ZPF}^2} S_{qq}[-\Omega]=(n+1)\Gamma_+[\Omega],\notag\\
&R_{\rm n \rightarrow n-1}[\Omega]=n\frac{g_a^2}{q_{\rm ZPF}^2} S_{qq}[\Omega]=n \Gamma_-[\Omega].
\label{R+-}
\end{align}
According to Eq.~(\ref{Cqqopto}), the photon emission (absorption) coefficient $\Gamma_{+(-)}$ can be expressed with the effective temperature and spectral density as
\begin{align}
&\Gamma_+[\Omega]=\frac{g_a^2}{q_{\rm ZPF}^2} S_{qq}[-\Omega]= 4 g_a^2 M \omega_m J_{\rm eff}[\Omega] \frac{1}{e^{\hbar \Omega\beta_{\rm eff}[\Omega]}-1},\notag\\
&\Gamma_-[\Omega]= \frac{g_a^2}{q_{\rm ZPF}^2} S_{qq}[\Omega]=4 g_a^2 M \omega_m J_{\rm eff}[\Omega]\frac{e^{\hbar \Omega\beta_{\rm eff}[\Omega]}}{e^{\hbar \Omega\beta_{\rm eff}[\Omega]}-1}.
\label{Gamma+-}
\end{align}
At equilibrium, the photon occupation number should satisfy the detailed-balance condition according to the ratio
\begin{align}
\frac{\bar{n}[\Omega]+1}{\bar{n}[\Omega]}=\frac{\Gamma_-[\Omega]}{\Gamma_+[\Omega]}=e^{\hbar\Omega \beta_{\rm eff}[\Omega]}=e^{\hbar(\omega_a-\nu_b) \beta_{\rm eff}[\omega_a-\nu_b]}.
\label{Detailed}
\end{align}
Note that since we are working in the rotating frame, the detailed-balance condition leads to the grand canonical distribution of photons associated with a frequency-dependent effective temperature, as predicted. Here $\hbar \nu_b$ takes the role of an effective chemical potential set by the driving frequency on the auxiliary beam-splitter mode $\hb$.

The effective spectral density that determines the coupling strength and the actual value of transition rates also matters. So far we have been working in the perfect cavity limit and neglecting the internal loss of the cavity. If the thermalization rates of the effective bath $R_{\rm n \rightarrow n+1}[\Omega]$ and $R_{\rm n \rightarrow n-1}[\Omega]$ are too slow such that one can no longer ignore the small cavity loss, the equilibrium condition becomes
\begin{align}
\frac{\bar{n}[\Omega]+1}{\bar{n}[\Omega]}=\frac{\Gamma_-[\Omega]+\kappa_a}{\Gamma_+[\Omega]}.
\label{Detailedwithloss}
\end{align}
The finite loss effect of the cavity will eventually destroy the desired grand canonical distribution. A strong enough coupling (determined by $J_{\rm eff}$) and enhanced coupling $g_a$ (determined by the power of the driving field $\hb_{\rm in}$) within the photonic bandwidth is therefore required to achieve efficient thermalization towards the grand canonical distribution.

To generate the equilibrium photonic state of interest, we aim at a well-defined (frequency-independent) bath temperature $T_B$ within the operating frequency bandwidth of interest $0 \leq \omega \lesssim |\Delta_a| \ll \omega_a$ and study the bath property within that range. We note again here that it is crucial to include the cooling mode $\hc$ to broaden the resonator linewidth and provide a lower effective temperature.

\section{Laser-Cooling-Dominated Limit}
\begin{figure}[htbp]
\begin{center}
\includegraphics[width=0.47 \textwidth]{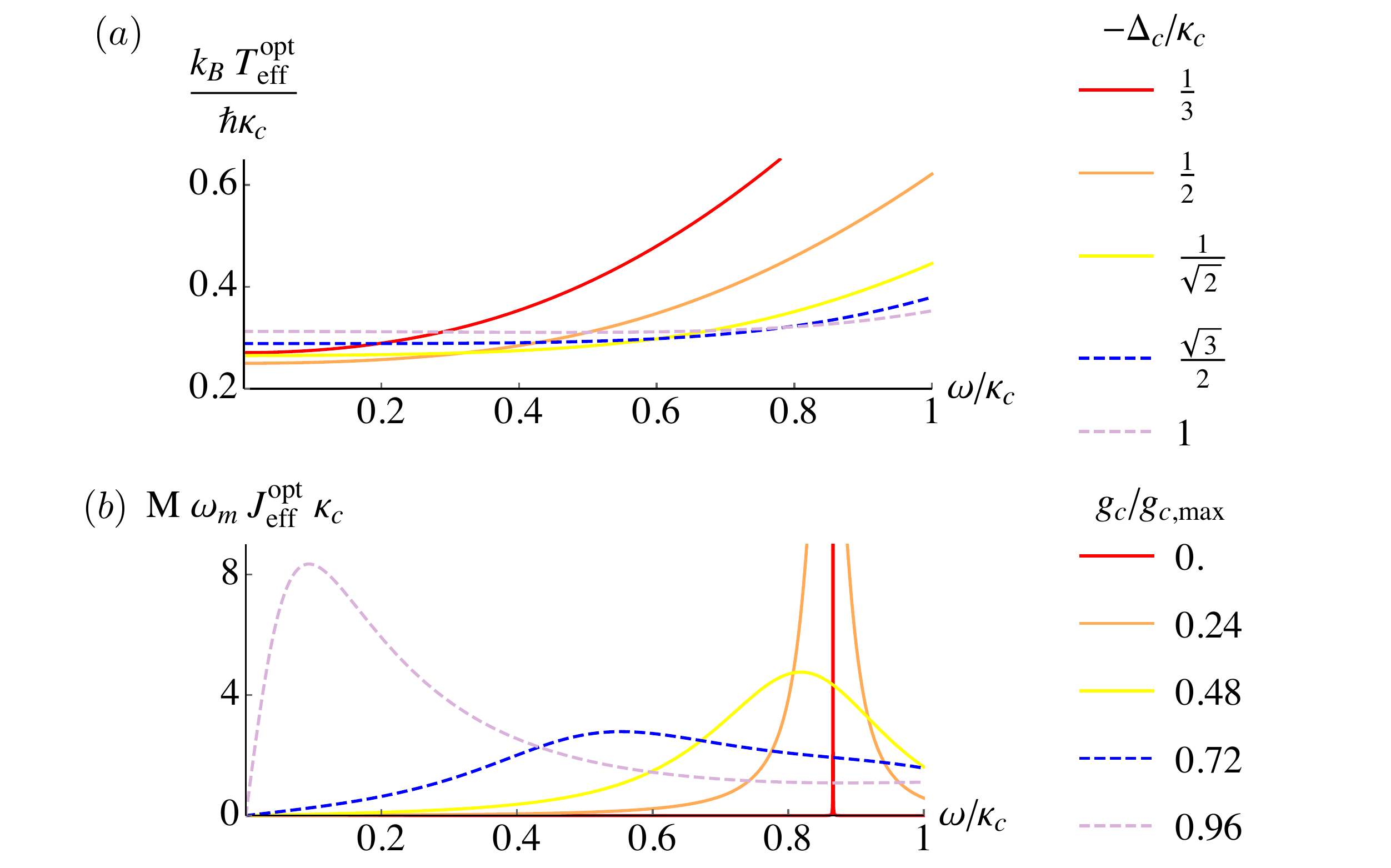}
\caption{(a) Effective temperature of the laser-cooled mechanical resonator as a bath under laser-cooling-dominated limit. (b) Effective spectral density of the bath under laser-cooling-dominated limit with the parameters $-\Delta_c=\omega_m=\sqrt{3}\kappa_c/2$ and $g_{\rm c,max}=\kappa_c/2$, and we have included $\beta=10^{-4}\omega_m$ and $\gamma_m=10^{-6}\omega_m$ for the $g_c=0$ case. Note that there are out-of-range peaks centered around the mechanical resonance $\omega_m$ in (b) for $g_c/g_{\rm c,max}=0, 0.24$.}
\label{fig:ToptJopt}
\end{center}
\end{figure}

First we look for an idealized case under the laser-cooling-dominated limit, $\gamma_m \approx 0$, such that one can achieve the minimum effective temperature by omitting the mechanical thermal environment and consider the laser cooling effect only. We then study the minimum effective temperature $T^{\rm opt}_{\rm eff}[\omega]$ and spectral density $J^{\rm opt}_{\rm eff}[\omega]$ under this limit, especially at low frequencies $\omega \ll |\Delta_c|, \kappa_c$ (see Fig.~\ref{fig:ToptJopt}). Note that we are working with the red-detuned regime $\Delta_c < 0$ for cooling process.
Specifically, according to the detailed-balance-like condition (\ref{detailbalance}),
\begin{align}
e^{\hbar \omega \beta^{\rm opt}_{\rm eff}[\omega]} = \frac{\pi\hbar G_c^2 L[\omega] }{\pi\hbar G_c^2 L[-\omega] }=\frac{(\omega-\Delta_c)^2+\kappa_c^2/4}{(\omega+\Delta_c)^2+\kappa_c^2/4}.
\label{detailbalancegsc}
\end{align}

Expanding this equation around $\omega=0$, we have
\begin{align}
\hbar \beta^{\rm opt}_{\rm eff}[\omega]=\frac{-4 \Delta_c}{\Delta^2_b+\kappa_c^2/4}-\frac{\Delta_c(4 \Delta_c^2-3\kappa_c^2)}{3(\Delta_c^2+\kappa_c^2/4)^3}\omega^2+{O}(\omega^4).
\label{dbgscexpand}
\end{align}
Thus we have an optimal choice of detuning $4\Delta_c^2=3\kappa_c^2$ to make the $\omega^2$ coefficient vanish. Note that Eq.~(\ref{dbgscexpand}) corresponds to a positive temperature for $\Delta_c<0$ associated with the cooling process. For a blue-detuned laser, the effective temperature is negative, representing a gain in the optomechanical system.

At low frequencies, the effective spectral density under the laser-cooling-dominated limit is
\begin{align}
\lim_{\omega \rightarrow 0}J^{\rm opt}_{\rm eff}[\omega]= \frac{-4 g_c^2 \Delta_c \kappa_c}{M \omega_m [\omega_m (\Delta_c^2+\kappa_c^2/4)+4 g_c^2 \Delta_c]^2}\omega=\eta^{\rm opt}_{\rm eff} \omega.
\label{Jeffgsclow}
\end{align}
We have introduced $g_c=G_c q_{\rm ZPF}=G_{c0} c_s q_{\rm ZPF}$ to express the pump-enhanced coupling strength as a frequency. Recall that the effective spectral density controls the induced damping rate by defining the effective damping kernel for the many-body system $\gamma_{\rm eff}(t)=\Theta(t)\frac{2}{\pi}\int_0^{\infty}\frac{J_{\rm eff}[\omega]}{\omega}\cos(\omega t) \mathrm{d}\omega.$ The effective spectral density is Ohmic in this regime, leading to an effective damping kernel $\gamma_{\rm eff}^{\rm opt}(t)=2 \eta^{\rm opt}_{\rm eff} \delta(t)$ corresponding to a memoryless Markovian-like damping term $-4 M \omega_m g_a^2 \eta_{\rm eff}^{\rm opt} \delta \dot{\hat{X}}_c$ in Eq.~(\ref{clangevin}) at low frequencies. Here $g_a= G_{a0} b_s q_{\rm ZPF}$ is the pump-enhanced coupling between $\hq$, $\ha$, and $\hb$ in the unit of frequency.

Note that for the red-detuned regime $\Delta_c<0$, $\eta^{\rm opt}_{\rm eff}$ diverges when $\omega_m (\Delta_c^2+\kappa_c^2/4)+4 g_c^2 \Delta_c$ approaches zero, which defines a critical value $g_{\rm c,max}^2 = \omega_m (\Delta_c^2+\kappa_c^2/4)/(4\abs{\Delta_c})$. An arbitrarily strong cooling rate can be achieved by increasing $g_c \lesssim g_{\rm c,max}$ towards the critical value. On the other hand, when one drives the cavity-enhanced coupling $g_c$ above this threshold value, the form of the low-frequency spectral density suggests a negative damping rate and the system is no longer stable. We will explore the detailed stability criteria for the system in the next section.

We remark that the initial mechanical thermal bath does not contribute to these equations as we are working in the laser-cooling-dominated limit. In reality, the resonator is always coupled to some thermal environment with finite dissipation $\gamma_m$ and we have seen that one cannot drive the laser intensity all the way to infinity before reaching dynamical instability. We are going to examine the maximum $g_c$ that ensures a stable perturbation around the steady-state solutions and then revisit the properties of the effective spectrum and temperature for a system with finite $\gamma_m$.

\section{Stability Criteria}
When the pump intensity driving the photonic mode is too strong, an optomechanical system may no longer be stable \cite{aspelmeyer2014}.
Here we study the stability criteria for the coupled linearized Heisenberg-Langevin equations (\ref{InOut}) in order to find the maximum pump intensities, or equivalently the maximum pump enhanced optomechanical coupling strengths, such that the steady-state solutions are stable and expansions around those solutions are still valid.

First we solve for the stability condition for the $
\hb$ and $\hq$ modes only before the interaction with the many-body system $\ha$ turns on.
We introduce dimensionless quadratures $\hx_c=\frac{1}{\sqrt{2}}(\dc + \dcd)$, $\hy_c=\frac{i}{\sqrt{2}}(\dcd - \dc)$, $\hQ=\sqrt{\frac{M \omega_m}{\hbar}} \hq$, $\hP=\sqrt{\frac{1}{\hbar M\omega_m}} \hp$, and the normalized stochastic force $\hat{\xi}(t)=\frac{\hat{F}_{\rm in}(t)}{\sqrt{\hbar M \omega_m}}$. The equations of motions becomes
\begin{align}
&\dot{\hat{X}}_c=-\frac{\kappa_c}{2} \hx_c-\Delta_c \hy_c+\sqrt{\kappa_c} \hx_{c,\rm in}(t),\\
&\dot{\hat{Y}}_c=\Delta_c \hx_c-\frac{\kappa_c}{2} \hy_c+2g_c \hQ+\sqrt{\kappa_c} \hy_{c,\rm in}(t),\\
&\dot{\hat{Q}}=\omega_m \hP,\\
&\dot{\hat{P}}=-\omega_m \hQ-\gamma_m \hP+2g_c \hx_c+\hat{\xi}(t),
\label{dimensionlessEOM}
\end{align}
corresponding to a matrix form
\begin{align}
\begin{pmatrix}
\dot{\hat{Q}}(t)\\ 
\dot{\hat{P}}(t)\\ 
\dot{\hat{X}}_c(t)\\
\dot{\hat{Y}}_c(t)
 \end{pmatrix}
= &
\begin{pmatrix}
0 & \omega_m & 0 & 0\\ 
-\omega_m & -\gamma_m & 2g_c & 0\\ 
0 & 0 & -\frac{\kappa_c}{2} & -\Delta_c\\
2g_c & 0 & \Delta_c & -\frac{\kappa_c}{2}
\end{pmatrix}
\begin{pmatrix}
\hQ(t)\\ 
\hP(t)\\ 
\hx_c(t)\\
\hy_c(t)
 \end{pmatrix}
 \notag\\ &
 +
 \begin{pmatrix}
0\\ 
\hat{\xi}(t)\\ 
\sqrt{\kappa_c} \hx_{\rm c,in}(t)\\
\sqrt{\kappa_c} \hy_{\rm c,in}(t)
 \end{pmatrix}.
\label{matrixEOM}
\end{align}
According to the Routh-Hurwitz criterion, the stability condition for the red-detuned cooling pump $\Delta_c<0$ is 
\begin{align}
4 g_c^2  \Delta_c+(\Delta_c^2+\kappa_c^2/4)\omega_m>0.
\label{RouthHurwitz}
\end{align}
Applying the optimal detuning for a fixed effective temperature around low frequencies $-\Delta_c=\sqrt{3}\kappa_c/2$, the condition sets an upper bound for the optical enhanced coupling $ g_c^2 \lesssim \frac{1}{2\sqrt{3}}\kappa_c \omega_m$. Recall that $g_c=G_{c0} c_s q_{\rm ZPF}$, $c_s=\sqrt{\kappa_c} c_{\rm in,s}/(i\Delta_c-\kappa_c/2)$, and the criterion sets the maximum intensity for the driving field $c_{\rm in}$.

We then further examine the stability condition when the coupling $G_{a0}$ is turned on. We omit $H_{a0}$ here and express the coupling in terms of the optical enhanced frequency $g_a= G_{a0} b_s q_{\rm ZPF}$, where $g_a$ is limited by buckling phase transitions \cite{xu2017}:
\begin{align}
\begin{pmatrix}
\dot{\hat{Q}}(t)\\ 
\dot{\hat{P}}(t)\\ 
\dot{\hat{X}}_c(t)\\
\dot{\hat{Y}}_c(t)\\
\dot{\hat{X}}_a(t)\\
\dot{\hat{Y}}_a(t)
\end{pmatrix}
=&
\begin{pmatrix}
0 & \omega_m & 0 & 0 & 0 & 0\\ 
-\omega_m & -\gamma_m & 2g_c & 0 & 2g_a & 0\\ 
0 & 0 & -\frac{\kappa_c}{2} & -\Delta_c & 0 & 0\\
2g_c & 0 & \Delta_c & -\frac{\kappa_c}{2} & 0 & 0\\
0 & 0 & 0 & 0& 0 & -\Delta_a\\
2g_a & 0 & 0 & 0 & \Delta_a & 0
\end{pmatrix}
\notag\\ &
\times
\begin{pmatrix}
\hQ(t)\\ 
\hP(t)\\ 
\hx_c(t)\\
\hy_c(t)\\
\hx_a(t)\\
\hy_a(t)
\end{pmatrix}
+
\begin{pmatrix}
0\\ 
\hat{\xi}(t)\\ 
\sqrt{\kappa_c}\hx_{c,\rm in}(t)\\
\sqrt{\kappa_c}\hy_{c,\rm in}(t)\\
0\\ 
0
 \end{pmatrix}.
\label{matrixEOMfull}
\end{align}

Under optimal detuning $-\Delta_c=\sqrt{3}\kappa_c/2$, and taking the limit $\gamma_m = 0$ since additional decay only enhances stability, the nontrivial stability conditions now read
\begin{align}
s_1&=\omega_m \kappa_c > 2 \sqrt{3} g_c^2, \notag\\
s_2&=-\Delta_a >0, \notag\\
s_3&= 2 \sqrt{3} \Delta_a g_c^2 - 4 g_a^2 \kappa_c - \Delta_a \kappa_c \omega_m>0.
\label{Bifurcation}
\end{align}
The condition $s_1$ is the same condition as before the interaction $g_a$ turned on. A negative detuning is necessary as suggested by $s_2$, and $s_3$ requires $\Delta_a <0$ and sets a more stringent upper bound for $g_c$, $\omega_m \kappa_c > 2 \sqrt{3} g_c^2 + \frac{4 g_a^2 \kappa_c}{\abs{\Delta_a}}$. Considering a weak system-bath coupling $g_a$ such that $\frac{4 g_a^2}{|\Delta_a| \omega_m} \ll 1$, one recovers the upper bound $g_{c,max}^2 \lesssim \frac{1}{2 \sqrt{3}} \kappa_c \omega_m$. 

\section{Beyond the Laser-Cooling-Dominated Limit}
We now revisit the effective low-frequency temperature and spectrum to include corrections from the mechanical dissipative environment. With finite $\gamma_m$ and $G_c$, according to Eq.~(\ref{detailbalance}), the effective inverse temperature at low frequencies is
\begin{align}
\lim_{\omega \rightarrow 0}\hbar \beta_{\rm eff}[\omega]= \frac{\hbar\beta [\gamma_m (\Delta_c^2 + \kappa_c^2/4)^2 - 4g_c^2 \Delta_c \kappa_c \omega_m]}{(\Delta_c^2+\kappa_c^2/4)[\gamma_m(\Delta_c^2+\kappa_c^2/4)+g_c^2 \beta \hbar \kappa_c \omega_m]}.
\label{betaeff}
\end{align}

To approach the laser-cooling-dominated limit, the conditions are $\gamma_m (\Delta_c^2 + \kappa_c^2/4)^2 \ll 4g_c^2 |\Delta_c| \kappa_c \omega_m$ and $\gamma_m(\Delta_c^2+\kappa_c^2/4) \ll g_c^2 \beta \hbar \kappa_c \omega_m$. Taking $g_c$ such that $g_{c,max}^2 \lesssim \frac{1}{2\sqrt{3}}\kappa_c \omega_m$ with the optimal detuning, the conditions become $ Q_m \gg  \kappa_c / \omega_m$ and $\frac{4}{\sqrt{3}} \pi Q_m f_m \gg k_B T/\hbar$; these conditions are satisfied for a high-$Q$ resonator in a quantum optomechanical regime.

Note that the optimal laser-cooling-dominated limit is achieved in the regime $\omega_m \approx \kappa_c$ in contrast to the usual side-band resolved cooling limit $\omega_m \gg \kappa_c$ to achieve the quantum ground state of the resonator. The different choice here is due to the fact that, while in the usual cooling process one hopes to have a narrow spectrum around the mechanical side band for efficient cooling, instead we are taking the low-frequency part of the resonator as a bath and thus require a larger linewidth $\kappa_c \sim \omega_m$ to broaden the noise spectrum from its center $
\omega_m$ towards low frequencies $\omega \sim 0$.

Around the laser-cooling-dominated limit, the correction to the temperature due to the mechanical environment is
\begin{align}
\lim_{\omega \rightarrow 0}\hbar \beta_{\rm eff}[\omega] &\approx -\frac{4 \Delta_c}{\Delta^2_b+\kappa_c^2/4} + \frac{4 \Delta_c+\hbar\beta(\Delta_c^2+\kappa_c^2/4)}{g_c^2 \hbar\beta \kappa_c \omega_m}\gamma_m \notag \\
&\quad +{O}(\gamma_m^2).
\label{betaeffgamma}
\end{align}
With this thermal correction, the effective temperature increases since $\hbar\beta(\Delta_c^2+\kappa_c^2/4)/\abs{\Delta_c}$ is typically small.

The effective spectral density including the thermal environment is
\begin{align}
\lim_{\omega \rightarrow 0}J_{\rm eff}[\omega] = \frac{\gamma_m (\Delta_c^2+\kappa_c^2/4)^2 -4 g_c^2 \Delta_c \kappa_c \omega_m}{M \omega_m (\omega_m (\Delta_c^2+\kappa_c^2/4)+4 g_c^2 \Delta_c)^2}\omega=\eta_{\rm eff} \omega.
\label{jeffgamma}
\end{align}
Note that this expression is exact without expansions in $\gamma_m$. At low frequencies the noise spectrum still behaves as an Ohmic heat bath and $\gamma_m$ contributes an extra damping rate to the system.

\section{Physical Design}

\begin{figure}[htbp]
\begin{center}
\includegraphics[width=0.4 \textwidth]{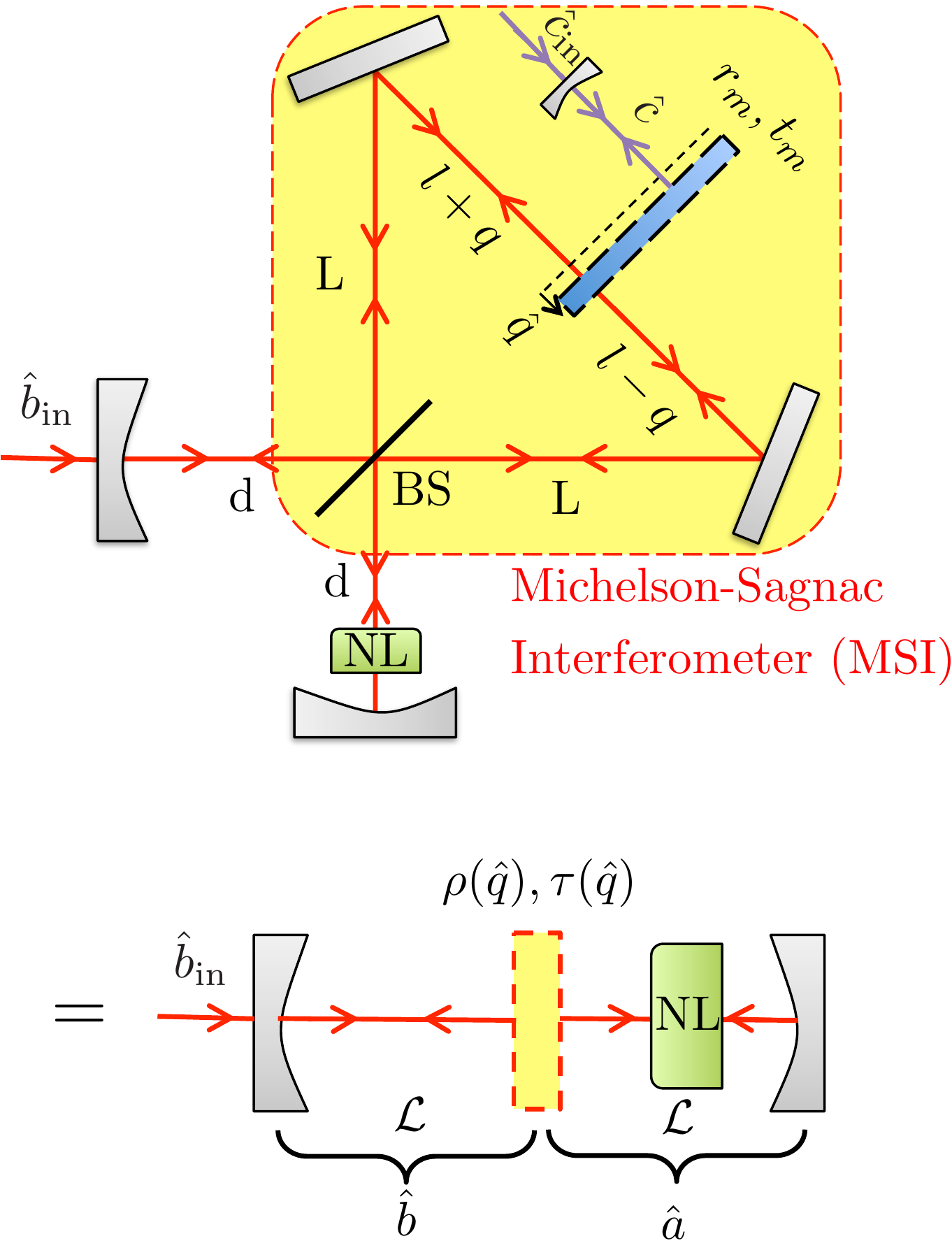}
\caption{Physical implementation that includes both pump rejection (via two red pathways of a Michelson-Sagnac interferometer) and reasonable optomechanical coupling even for small reflectivity mirrors, such as thin SiN membranes. The equivalent two cavity modes showing the nonlinear (marked NL) media that forms part of the quantum simulator and the pumped mode are shown in the lower section. Another spatially independent cavity $\hc$ (purple pathway), which takes the resonator as one of the end mirrors and is driven by a red-detuned laser field $\hc_{\rm in}$, is added to laser cool the resonator.}
\label{fig:MSI}
\end{center}
\end{figure}

With our theoretical analysis in place, we look for a potential physical design to realize our optomechanical implementation. One essential component of the theory is the purely beam-splitter-type coupling between our system $\ha$, the mechanical resonator $\hq$, and the beam-splitter auxiliary optical mode $\hb$. To achieve a near-equilibrium grand canonical ensemble of photons, we require the drive to enter as a classical amplitude of $\hb$ mediating the sinusoidal parametric coupling, without driving the many-body system ($\ha$) directly. While a beam-splitter-type interaction is common in so-called mirror-in-the-middle systems \cite{jayich2008,xu2014,stambaugh2015}, the strong driving field on $\hb$ inevitably leaks into the system $\ha$ through the translucent middle mirror and generates unwanted steady-state coherence in $\ha$ in the simple mirror-in-the-middle geometry.

Here we suggest a potential experimental design via a Michaelson-Sagnac interferometer (MSI). This topology was first proposed to apply power and signal recycling techniques on translucent membrane resonators for accessing a quantum radiation pressure noise regime \cite{yamamoto2010} and can be used to realize generalized optomechanical coupling and cooling \cite{sawadsky2015}. In the interferometer geometry (see Fig.~\ref{fig:MSI}), one uses a translucent (with reflectivity $r_m$ and transmissivity $t_m$) subwavelength mechanical resonator (for example, thin SiN membranes) as a common end mirror for the two arms of the Michelson interferometer, while the transmitted light through the resonator forms a Sagnac mode. Note that the Sagnac mode is insensitive to the resonator position. This set up is equivalent to placing a fixed mirror at equal lengths between two high quality factor cavity end mirrors while the effective reflectivity $\rho$ and transmissivity $\tau$ of the middle mirror change with the resonator motion $\hat{q}$.

For pump rejection, we operate the inteferometer at its dark fringe condition: At the equilibrium position of the resonator, the Michelson and Sagnac modes form a total destructive interference at the output end, with $\abs{\rho(\hat{q}=0)}=1$ and $\tau(\hat{q}=0)=0$. The two cavity modes are initially decoupled for zero displacement of the mechanical resonator. Using the transfer matrix method, following the supplemental material of Ref.~\cite{sawadsky2015}, and applying the boundary condition at the two cavity end mirrors to solve for normal mode eigenvalues \cite{jayich2008}, we find that the Hamiltonian of the composite system reads
\begin{align}
\hat{H}_{\rm MSI}&=\omega_0 \had\ha+\hbar\omega_0 \hbd \hb+\frac{\hp^2}{2M}+\frac{M\omega_m^2 \hq^2}{2}\notag \\
&\quad + \frac{r_m \omega_0}{\mathcal{L}}\hq(\had \hb+ \hbd \ha),
\label{MSI}
\end{align}
Here $\omega_a=\omega_b=\omega_0$ is the resonant frequency of the cavity modes $\ha$ and $\hb$, $r_m$ is the complex reflectivity of the mechanical resonator, and $\mathcal{L}=d+L+l$ is the effective cavity length.

Thus we arrive at a purely beam-splitter-type coupling between the mechanical system and the cavity modes. In the symmetric MSI geometry at the dark fringe condition, the driving field on $\hb$ merely mediates the parametric coupling without pumping the many-body photonic system $\ha$ directly. One can include an additional laser cooling mode $\hc$, for example, by adding another spatially independent cavity to complete the bath engineering story as described in Fig.~\ref{fig:Schematic}. This concept of using purely beam-splitter optomechanical interactions for pump rejection should be extensible to many other configurations.

\section{Outlook}
Our approach for controlling the chemical potential and temperature of light suggests a path forward for creating equilibrium many-body states of photon-based quantum simulators.
However, key questions remain, including the best way to create nonlinear optical or microwave terms
as well as methods in the microwave domain for determining the photonic statistics to confirm our grand canonical ensemble prediction. Furthermore,
intriguing new challenges await, particularly with regard to other conserved quantities and their associated thermodynamic Lagrange multipliers. The general approach used here may be extensible to other such scenarios, which may allow for the exploration of a wide range of thermodynamic ensembles. We also note that our technical implementation may have the many-mode extension necessary to generate thermodynamic equilibrium in a macroscopic system, which is left for future work.

\begin{acknowledgments}
We thank T. P. Purdy, M. Hafezi, S. Girvin, G. Zhu, J. Stirling, M. P. Zwolak, E. C. Benck, X. Xu and S. Ragole for helpful discussions. Funding was provided by Physics Frontier Center at the JQI.
\end{acknowledgments}



\bibliographystyle{apsrev4-1} 
\bibliography{PaperChemOpto}

\end{document}